# Causality-based CTR Prediction using Graph Neural Networks

Panyu Zhai[a], Yanwu Yang[a] and Chunjie Zhang[b]

[a]School of Management, Huazhong University of Science and Technology, Wuhan 43004, China, {zhaipanyu.isec, yangyanwu.isec}@gmail.com

[b]Institute of Information Science, Beijing Jiaotong University, Beijing 100044, China, cjzhang@bjtu.edu.cn

**Abstract:** As a prevalent problem in online advertising, CTR prediction has attracted plentiful attention from both academia and industry. Recent studies have been reported to establish CTR prediction models in the graph neural networks (GNNs) framework. However, most of GNNs-based models handle feature interactions in a complete graph, while ignoring causal relationships among features, which results in a huge drop in the performance on out-of-distribution data. This paper is dedicated to developing a causality-based CTR prediction model in the GNNs framework (Causal-GNN) integrating representations of feature graph, user graph and ad graph in the context of online advertising. In our model, a structured representation learning method (GraphFwFM) is designed to capture high-order representations on feature graph based on causal discovery among field features in gated graph neural networks (GGNNs), and GraphSAGE is employed to obtain graph representations of users and ads. Experiments conducted on three public datasets demonstrate the superiority of Causal-GNN in AUC and Logloss and the effectiveness of GraphFwFM in capturing high-order representations on causal feature graph.

# 1. Introduction

A variety of advertising forms have emerged with the development of the Internet. Online advertising and recommender systems have become dominant channels to promote products or services for worldwide firms (Yang et al., 2017). In 2020, online advertising amounted to $378.16 billion, in spite of the negative economic impact of the COVID-19 pandemic (Statista, 2022). According to IAB report (IAB, 2022), the online advertising revenue soared 35% to $189 billion in 2021 in U.S. alone, which is the highest growth since 2006. As recognized by both researchers and practitioners, predicting the click-through rate (CTR) is a critical issue in the multi-billion-dollar online advertising industry (McMahan et al., 2013).

As a prevalent problem in online advertising, CTR prediction has attracted plentiful research efforts from academia and industry. Existing CTR prediction research mainly focuses on feature interactions based on factorization machines (FMs), deep neural networks (DNNs) and graph neural networks (GNNs) (Yang & Zhai, 2022). Although FMs theoretically support high-order representations, FMs-based models typically use pairwise feature interactions for CTR prediction due to the complexity raised by high-order interactions (Rendle, 2010; Li et al., 2021b). DNNs have been integrated with LR and FMs into a rich set of modeling frameworks to capture sophisticated feature interactions, such as Wide & Deep (Cheng et al., 2016), NFM (He & Chua, 2017), DeepFM (Guo et al., 2017). Although DNN-based models are generally better than FMs attributing to their deeper networks, interactions are performed in implicit manners and thus with low interpretability. Moreover, DNN- and FMs-based models explore feature interactions in Euclidean spaces, which may be unrealistic in most scenarios. In contrast, GNNs-based models investigate feature interactions in non-Euclidean spaces, by converting feature interactions to node interactions in a graph structure. In GNNs, features (nodes) aggregate information from their neighbors to update their hidden states (Song et al., 2021).

Recently, quite a few studies have been reported to establish CTR prediction models in the GNNs framework, by modeling high-order interactions in feature graph (Li et al., 2019b; Li et al., 2021a; Li et al., 2021b), designing graph intention networks (Li et al., 2019a) and dynamic sequential graph (Chu et al., 2021) to enrich users' behaviors, constructing attribution graph and collaborative graph to address the sparsity issue (Guo et al., 2021). However, most of existing GNNs-based models handle feature interactions by aggregating information from all neighbors equally in a complete graph (Tao et al., 2020). On one hand, these models are vulnerable to over-smoothing due to the complete-graph aggregation. On the other hand, these models fail to consider inner relationships among features and thus suffer from the performance drop on the out-of-distribution data (Wu et al., 2022). In effect, causal inference endows the prediction with better model generalization and performance robustness (Zhang et al., 2022a). Moreover, causality-based feature interactions can enhance the interpretability of aggregation functions in GNNs-based models in that causality characterizes inherent relationships among features (Sui et al., 2022). To the best of our knowledge, this is the first research designing causality-based CTR prediction models in the GNNs framework.

This paper proposes a causality-based CTR prediction model in the GNNs framework (Causal-GNN). In order to capture the fact that each feature may behave differently when interacting with others, we develop a structured representation learning approach (GraphFwFM) to capture feature representations based on causal discovery among field features in gated graph neural networks (GGNNs). We also construct user graph and ad graph through statistical analysis reflecting the similarity among users and among ads, respectively, and employ GraphSAGE to generate embeddings by sampling and aggregating from the local neighborhood in user graph and ad graph. The multi-head attention mechanism is utilized to fuse graph representations of features, users and ads, and the neural-based attention aware predictor takes attention-weighted representations to predict the clicking probability. We conducted experiments on three public datasets (i.e., Criteo, Avazu and MovieLens-1M) to evaluate the performance of Causal-GNN by comparing with several state-of-the-art baselines. Experimental results illustrate the superiority of Causal-GNN in AUC and Logloss and the effectiveness of GraphFwFM in modeling high-order representations in causal feature graph.

The main contribution of this study can be summarized as follows. First, we propose a causality-based CTR prediction model in the GNNs framework integrating multiple representations of graph enabled features. Second, we design a structured representation learning approach (GraphFwFM) to capture causal feature representations in GNNs. Third, experiments are conducted on three public datasets to demonstrate the superiority of the Causal-GNN model and the merit of GraphFwFM in CTR prediction tasks.

The remainder of this paper is organized as follows. Section 2 gives a brief review on the related work on CTR prediction and discusses the linkage of this research with the extant literature. Section 3 presents the modeling structure and details of Causal-GNN. Experimental evaluations are reported in Section 4, and we conclude this research in Section 5.

## 2. Related work

In this section, we first provide a brief review on four major classes of advertising CTR prediction models reported in the literature, then primarily focus on GNNs-based models for CTR prediction.

### 2.1 The classification of CTR prediction models

In the literature on advertising CTR prediction, researchers have primarily explored four classes of modeling frameworks including multivariate statistical models, factorization machines (FMs) based models, deep learning models and tree models. For a comprehensive survey on CTR prediction models in online advertising, refer to see Yang & Zhai (2022).

(1) **Multivariate statistical models** include logistic regression (LR) (Richardson et al., 2007) and Poly2 model (Chang et al., 2010), which independently uses an individual feature or simply considers interactions between features (Chang et al., 2010).

(2) **Factorization machines (FMs)-based models** (e.g., FMs, FFMs, FwFMs and AFM) effectively capture pairs of feature interactions by using the factorized mechanism (Rendle, 2010; Xiao et al., 2017). Formally, in the FMs-based modeling framework, the inner product is used to capture pairwise feature interactions of latent vectors between features. FMs-based models are efficient in achieving low-order feature interactions and show the recognized performance in CTR prediction tasks (Guo et al., 2017).

(3) **Deep learning models** are generally utilized to capture high-order feature interactions for CTR prediction, including standard long short-term memory (LSTM) (Hochreiter and Schmidhuber, 1997), convolutional neural network (CNN) (Zhang et al., 2022c), factorization machine supported neural network (FNN) (Zhang et al., 2016). In order to capture feature interactions of multiple orders flexibly, a bunch of ensemble models have been reported integrating deep learning models with (either low-order or high-order) explicit components, e.g., Wide & Deep (Cheng et al., 2016), DeepFM (Guo et al., 2018), Deep & Cross network (DCN) (Wang et al., 2017) and xDeepFM (Lian et al., 2018).

(4) **Tree models** are developed based on the idea of boosting in ensemble learning, including Gradient boosting decision tree (GBDT) (Friedman, 2001) and XGBoost (Chen & Guestrin, 2016), which have shown considerable successes in CTR prediction tasks (He et al., 2014). Tree model largely suffers from the sparsity problem. To this end, researchers have extensively explored ensemble models combining tree models with various modeling components, e.g., GBDT+gcForest (Qiu et al., 2018), GBDT+DNN (Ke et al., 2019), XGBoost+FwFMs (Shi et al., 2019) and XGBoost+DeepFM (An & Ren, 2020).

However, the above CTR prediction models focus on unstructured combinations of features and are largely limited by the prediction capability and/or implicit forms. Moreover, these models handle feature interactions in Euclidean spaces, which may be unrealistic in most scenarios.

## 2.2 GNNs in CTR prediction

Graph neural networks (GNNs) are an emerging modeling framework to broaden the feature horizon of CTR prediction in non-Euclidean spaces and support more interpretable models. Note that GNNs-based CTR prediction falls into the third class (i.e., deep learning models) as discussed in the previous section.

Due to their strength in graph representations, GNNs have been used to alleviate the feature sparsity and behavior sparsity problems in CTR prediction, by converting feature interactions into node interactions in a graph structure. Li et al. (2019b) proposed a feature interaction GNNs (Fi-GNN) where field-aware feature interactions were realized through assigning two interaction matrices to each node in a complete graph; in a later work, Li et al. (2021a) used the pre-trained GNNs to generate explicit semantic-cross features and applied the weighted square loss to compute the importance of these features.

Users' potential interests are beneficial to predicting their future behaviors. Graph models are efficient in representing users' rich, diverse and fluid interests (Li et al. 2019a; Wang et al., 2022) and latent correlations among various interests (Wang et al., 2021), which can be extracted from their past behaviors and interactions with items (Zhang et al., 2022b) by using hierarchical attention and multi-head attention mechanisms (Zheng et al., 2022), graph-masked transformer (Min et al., 2022), and triangle graph interest network (Jiang et al., 2022). Collaborative information between users and items is very valuable for extracting users' interests and in turn improving CTR prediction. In particular, collaborative graphs constructed based on users' behaviors and items' attributions can enhance feature and behavior embeddings in GNNs frameworks (Guo et al., 2021; Zhang et al., 2022b). Aiming to mine potential users' interests and real intentions from their behaviors, Min et al. (2022) constructed a heterogeneous graph with four types of interactions and used graph-masked transformer to capture highly representative embeddings of users and items, and Jiang et al. (2022) proposed a triangle graph interest network (TGIN) that utilized triangles in the neighborhood in the item co-occurrence graph to extract implicit user interests and aggregated the information of several interest units represented by triangles.

Online advertising is an extremely dynamic and complex environment for marketing and promotions (Yang et al., 2022; Li & Yang, 2022). In order to capture users' real-time interest, Li et al. (2019a) designed a Graph Intention Network (GIN) based on a co-occurrence commodity graph and adopt multi-layer graph diffusion to enrich users' behaviors, Chu et al. (2021) applied graph convolutional networks on dynamic sequential graphs of users and items to obtain representations of the target user and the candidate item iteratively, and Wang et al. (2022) proposed a dynamic graph-based disentangled representation framework (DisenCTR) where a disentangled representation learning component was used to extract users' diverse interests from a time-evolving user-item interaction graph.

Existing GNNs-based models can capture sophisticated feature interactions and representations of users' behaviors and interests helpful for CTR prediction. However, GNNs are vulnerable to data biases and especially shortcut features (Wu et al., 2022). That is, when the data is out-of-distribution, the performance of GNNs-based models may drop drastically. Moreover, most of existing GNNs-based models handle feature interactions in a complete graph, while ignoring inner relationships among features (e.g., causality). Theoretically, causality can improve the generalization of prediction models and boost the model performance on out-of-distribution data (Wu et al., 2022). It is worthwhile to note that causal inference provides an alternative perspective for the interpretability of aggregation functions in the GNNs framework. In addition, the input for existing CTR prediction models is generally restricted to field features and interactions among features. As a matter of fact, there exist various types of graph-enabled features (e.g., relationships between users and between advertisements) valuable for CTR predictions.

This research aims to promote the performance of CTR prediction by integrating causal feature representations in the GNNs modeling framework (Causal-GNN). Specifically, we use multiple GNNs to extract useful information from causal feature graph, user graph and ad graph constructed through causal inference and statistical analysis. Moreover, we develop a GGNNs-based representation learning approach (GraphFwFM) to capture high-order representations based on causal discovery among field features. In addition to causal feature graph, user graph and ad graph representation learning components encapsulated in our CTR prediction model are expected to address the sparsity problem. This is the first research using causal inference to facilitate CTR prediction in the GNNs framework integrating multiple graph representations.

## 3. The Model

In this section, we start with the modeling structure of our CTR prediction model (Causal-GNN), then turn to details of each component. Table 1 lists the notations used in this paper.

Table 1. Notations

| Terms | Definition |
|---|---|
| $f_k$ | The $k$-th field feature. |
| $u_i$ | The $i$-th user. |
| $a_j$ | The $j$-th ad. |
| $S$ | The number of field features. |
| $N$ | The number of users. |
| $M$ | The number of ads. |
| $\mathcal{G}_f(\mathcal{F}, \mathcal{E}_{\mathcal{F}})$ | Feature graph where $\mathcal{F}$ is the set of features and $\mathcal{E}_{\mathcal{F}}$ is the set of edges between features. |
| $\mathcal{G}_u(\mathcal{U}, \mathcal{E}_u)$ | User graph where $\mathcal{U}$ is the set of users and $\mathcal{E}_u$ is the set of edges between users. |
| $\mathcal{G}_a(\mathcal{A}, \mathcal{E}_a)$ | Ad graph where $\mathcal{A}$ is the set of ads and $\mathcal{E}_a$ is the set of edges between ads. |
| $w_{k'k}^{(f)}$ | The weight of an edge from $f_{k'}$ to $f_k$ in feature graph $\mathcal{G}_f$. |
| $w_{i'i}^{(u)}$ | The weight of an edge from $u_{i'}$ to $u_i$ in user graph $\mathcal{G}_u$. |
| $w_{j'j}^{(a)}$ | The weight of an edge from $a_{j'}$ to $a_j$ in ad graph $\mathcal{G}_a$. |
| $\tilde{e}_{f_k}$ | The field embedding vector of feature $f_k$. |
| $e_{f_k}$ | The graph embedding vector of feature $f_k$. |
| $e_{u_i}$ | The graph embedding vector of user $u_i$. |
| $e_{a_j}$ | The graph embedding vector of ad $a_j$. |
| $\mathbb{e}_{f_k}$ | The concatenation of field embedding ($\tilde{e}_{f_k}$) and graph embedding ($e_{f_k}$). |
| $H_{f_k}^{(l)}$ | The representation of feature $f_k$ in the $l$-th feature graph. |
| $H_{\mathcal{N}(f_k)}^{(l)}$ | The aggregated representation of feature $f_k$ in the $l$-th feature graph. |
| $H_{u_i}^{(l)}$ | The representation of user $u_i$ in the $l$-th user graph. |
| $H_{\mathcal{N}(u_i)}^{(l)}$ | The aggregated representation of user $u_i$ in the $l$-th user graph. |
| $H_{a_j}^{(l)}$ | The representation of ad $a_j$ in the $l$-th ad graph. |
| $H_{\mathcal{N}(a_j)}^{(l)}$ | The aggregated representation of ad $a_j$ in the $l$-th ad graph. |
| $\alpha_{k'k}$ | The importance of feature $f_{k'}$ in the graph aggregation for feature $f_k$. |
| $H_{f_k}$ | The representation of feature $f_k$ after the multi-head self-attention. |
| $H_{u_i}$ | The representation of user $u_i$ after the multi-head self-attention. |
| $H_{a_j}$ | The representation of ad $a_j$ after the multi-head self-attention. |
| $K_f$ | The depth of graph neural networks in feature graph. |
| $K_u$ | The depth of graph neural networks in user graph. |
| $K_a$ | The depth of graph neural networks in ad graph. |

## 3.1 Modeling Structure

We propose a CTR prediction model (i.e., Causal-GNN) including five major components, as shown in Figure 1. The first component is the graph construction layer, which builds feature graph, user graph and ad graph through causal learning and statistical analysis. The second component is the embedding layer, which encodes each field as a binary vector by one-hot encoders, and embeds field features as dense vectors through field-aware embedding and graphs as latent representations through graph embedding. The third component is the GNNs layer, which learns graph representations of features, users and ads by using GraphFwFM and GraphSAGE. The fourth is the attention layer, which processes the output of the GNNs with multi-head attention mechanisms, and the fifth, i.e., the prediction layer, utilizes a neural-based attention method to integrate representations of users, ads and features to conduct CTR prediction.

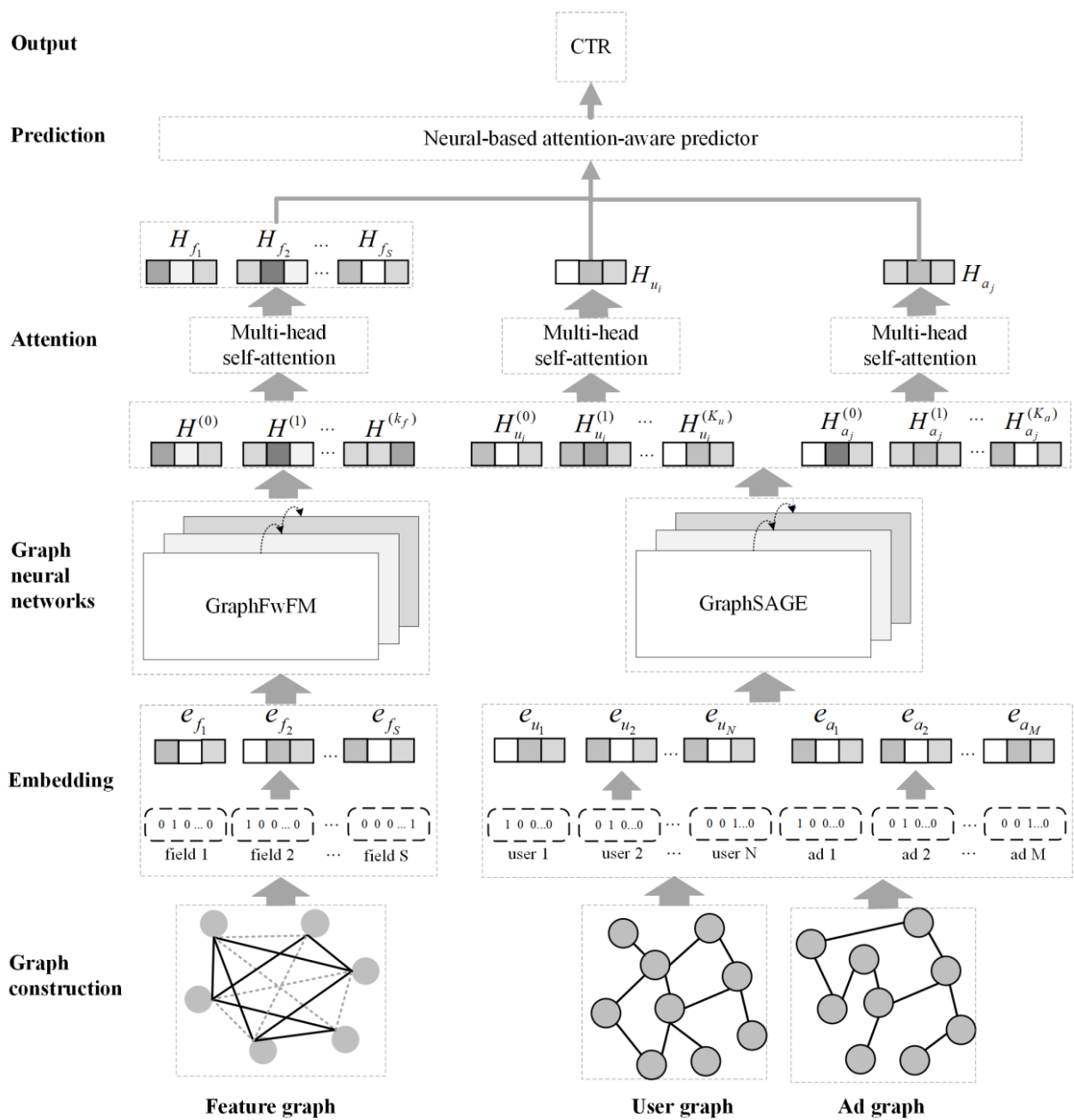

Figure 1. The structure of the proposed CTR prediction model (Causal-GNN)

## 3.2 Graph construction

The graph construction layer establishes three graphs namely feature graph, user graph and ad graph through causal learning and statistical analysis on historical advertising logs.

Conceptually, a causal graph is a directed graph with nodes denoting features (or variables) and edges denoting the dependencies among these features (Helmert, 2004; Hu et al., 2014; Kocaoglu et al., 2017). In our context, we construct a causal feature graph to capture high-order feature interactions for CTR prediction. Whereas there is no causal relationship among users and among ads, we construct user graph and ad graph based on the similarity.

### 3.2.1 Causal learning in feature graph

This research constructs a causal graph with field features of each instance through causal learning. In the causal feature graph $\mathcal{G}_f(\mathcal{F}, \mathcal{E}_\mathcal{F})$, each node corresponds to a field $f_k$, i.e., $\mathcal{F} = \{f_1, f_2, \ldots, f_S\}$, and an edge from one node to another is directed and denotes the causal relationship on the path. Causal relationships in $\mathcal{G}_f(\mathcal{F}, \mathcal{E}_\mathcal{F})$ are represented with a weighted adjacency matrix $W^{(f)} \in R^{S \times S}$.

In most of existing causality learning frameworks, directed acyclic graphs (DAGs) are formed to represent the causal structure of the feature space (Zheng et al., 2018; Wei et al., 2020). Given an independent and identically distributed (i.i.d.) sample $X$ ($X \in R^{S \times q}$, where $S$ the number of field features and $q$ are the dimension of a (vector) field, respectively, causal learning aims to recover a DAG structure over features (represented by $W^{(f)}$) from $X$. Following prior studies (Yu et al., 2019; Zhang et al., 2019), causal learning generalizes the SEM model in the GNNs framework, which is given as

$$f^{-1}(X) = W^{(f)^T} f^{-1}(X) + g(Z), \quad (1)$$

where $Z \in R^{S \times q}$ is the noise matrix, g and $f$ are parameterized functions on $Z$ and $X$, respectively.

Causal learning based on variational Bayes (Kingma & Welling, 2013) approximates the true posterior $p_\theta(Z|X)$ with the variational posterior $q_\phi(Z|X)$, by minimizing the Kullback-Leibler (KL) divergence of the latter from the former (Blundell et al., 2015), given as

$$\arg\min_{\theta,\phi} D_{KL}[q_\phi(\mathbf{Z}|\mathbf{X})||p_\theta(\mathbf{Z}|\mathbf{X})]$$

$$= \arg\min_{\theta,\phi} \int q_\phi(\mathbf{Z}|\mathbf{X}) \log \frac{q_\phi(\mathbf{Z}|\mathbf{X})}{p_\theta(\mathbf{Z})p_\theta(\mathbf{X}|\mathbf{Z})} d\mathbf{Z}$$

$$= \arg\min_{\theta,\phi} D_{KL}[q_\phi(\mathbf{Z}|\mathbf{X})||p_\theta(\mathbf{Z})] - E_{q_\phi(\mathbf{Z}|\mathbf{X})}[\log p_\theta(\mathbf{X}|\mathbf{Z})], \quad (2)$$

where $D_{KL}$ is the KL-divergence, $\theta = (M_X, S_X)$ and $\phi = (M_Z, S_Z)$ are generative parameters and variational parameters, respectively.

The resulting cost function in Equation (2) is known as the expected lower bound, whose negative form is called the variational lower bound or the evidence lower bound (ELBO). Given a distribution of $\mathbf{Z}$ and a set of samples $\mathbf{X}^1, \mathbf{X}^2, \ldots, \mathbf{X}^n$, the loss is defined as the mean negative lower bound, which is reformulated as

$$L_{ELBO} = -\frac{1}{n}\sum_{k=1}^{n} D_{KL}(q_\phi(\mathbf{Z}|\mathbf{X}^k)||p_\theta(\mathbf{Z})) - E_{q_\phi(\mathbf{Z}|\mathbf{X}^k)}[\log p_\theta(\mathbf{X}^k|\mathbf{Z})]. \quad (3)$$

Density functions $q_\phi(\mathbf{Z}|\mathbf{X}^k)$ and $p_\theta(\mathbf{X}^k|\mathbf{Z})$ can be obtained through probabilistic encoder and decoder of Bayesian neural networks (BNNs) (Kingma & Welling, 2013; Yu et al., 2019). Specifically, the encoder instantiates $f^{-1}$ with BNNs and g with an identity mapping to obtain parameters ($M_Z$ and $S_Z$) of the distribution of the variational posterior $q_\phi(\mathbf{Z}|\mathbf{X})$, and the decoder employs inverse functions of $f$ and g to obtain parameters ($M_X$ and $S_X$) of the distribution of the true posterior $p_\theta(\mathbf{X}|\mathbf{Z})$.

Taking into account the acyclicity constraint, causal learning can be transformed into the following optimization problem through the augmented Lagrangian method with $\mathcal{L}_1$-regularization (Ng et al., 2019), which is given as

$$(\mathbf{W}^{(f)}, \boldsymbol{\Theta}) = \arg\min_{\mathbf{W}^{(f)}, \boldsymbol{\Theta}} (-L_{ELBO} + \lambda ||\mathbf{W}^{(f)}||_1 + \alpha h(\mathbf{W}^{(f)}) + \frac{\rho}{2}|h(\mathbf{W}^{(f)})|^2), \quad (4)$$

$$s.t. \quad h(\mathbf{W}^{(f)}) = tr\left[\left(\mathbf{I} + \alpha \mathbf{W}^{(f)} \circ \mathbf{W}^{(f)}\right)^S\right] - S = 0, \quad (5)$$

where $tr$ is the trace of matrix $\left(\mathbf{I} + \alpha \mathbf{W}^{(f)} \circ \mathbf{W}^{(f)}\right)^S$, $\boldsymbol{\Theta}$ is a set of parameters of BNNs in variational autoencoders, $\alpha$ is the Lagrange multiplier and $\rho$ is the penalty parameter.

The optimization problem (Equations 4 and 5) can be solved through updating $\alpha$ and increasing $\rho$ by using stochastical optimization solvers to obtain the adjacency matrix $\mathbf{W}^{(f)}$ and $\boldsymbol{\Theta}$. In this research, causal relationships described by $\mathbf{W}^{(f)}$ is used to feed the causal feature graph representation learning discussed in Section 3.4.1.

### 3.2.2 User graph and ad graph construction

User graph is established based on common ads displayed to each pair of users. In the user graph $\mathcal{G}_u(\mathcal{U}, \mathcal{E}_u)$, $\mathcal{U} = \{u_1, u_2, \ldots, u_N\}$, each user corresponding to a node ($u_i$); there exists an edge from $u_{i'}$ to $u_i$ if one or more common ads are displayed to the two users. The weight of the edge from $u_{i'}$ to $u_i$ is calculated as $w_{i'i}^{(u)} = \frac{|S_{i'} \cap S_i|}{|S_i|}$, where $S_{i'}, S_i \in \mathcal{A}$ denote the collections of advertisements displayed to $u_{i'}$ and $u_i$, respectively.

Ad graph is constructed in a similar way. That is, in the ad graph $\mathcal{G}_a(\mathcal{A}, \mathcal{E}_a)$, $\mathcal{A} = \{a_1, a_2, \ldots, a_M\}$, each node corresponds to an advertisement ($a_j$) and the presence of an edge means that the two ads are displayed to one or more common users. The weight of the edge from $a_{j'}$ to $a_j$ is calculated as $w_{j'j}^{(u)} = \frac{|A_{j'} \cap A_j|}{|A_j|}$, where $A_{j'}, A_j \in \mathcal{U}$ denote the sets of users to whom $a_{j'}$ and $a_j$ are exposed, respectively.

## 3.3 Embedding

In this Section, we present embedding techniques to transform features, users and advertisements into dense vectors.

### 3.3.1 Field feature embedding

In CTR prediction, multi-field categorical features are usually transformed into a vector containing 0 and 1 through one-hot encoding. For example, 'ad position=3' can be encoded as '[0, 0, 1,…, 0]', which are high-dimensional sparse vectors. Hence, it is necessary to transform one-hot vectors $x_{f_k}$ ($f_k \in \mathcal{F}$) into low-dimensional dense feature vectors through field-aware embedding methods (Ma et al., 2016; Wang et al., 2017; Xie et al., 2019). Formally, each field $f_k$ can be represented as a dense vector as follows.

$$\tilde{e}_{f_k} = W_{emb} x_{f_k}, \quad (6)$$

where $W_{emb} \in \mathbf{R}^{d \times n_c}$ is the embedding matrix, $n_c$ is the number of features, $d$ is the embedding size, $\tilde{e}_{f_k} \in \mathbf{R}^d$ is the embedding vector of field feature $f_k$.

### 3.3.2 Graph embedding

This research employs DeepWalk capture relationships in feature graph $\mathcal{G}_f(\mathcal{F}, \mathcal{E}_f)$, user graph $\mathcal{G}_u(\mathcal{U}, \mathcal{E}_u)$ and ad graph $\mathcal{G}_a(\mathcal{A}, \mathcal{E}_a)$, which is a prevalent graph embedding technique. DeepWalk treats random walks in a graph as a sentence to learn the embedding of each node (Perozzi et al., 2014; Goyal & Ferrara, 2018).

In a directed graph (e.g., feature graph, user graph and ad graph), a random walk $\mathcal{W}_i$ rooted at $node_i$ with length $t$ can be made by the RandWalk algorithm, i.e., $\mathcal{W}_i = RandWalk(G, node_i, t)$. Then the embedding of each user can be learnt with the mapping function $\boldsymbol{E} \in \boldsymbol{R}^{N \times d}$ using the SkipGram algorithm (Perozzi et al., 2014) with a specific window size $w$, which is given as

$$\boldsymbol{e} = SkipGram(\boldsymbol{E}, \mathcal{W}_i, w), \quad (7)$$

In Equation (7), $\boldsymbol{e}$ is instantiated as $\boldsymbol{e}_f \in \boldsymbol{R}^{N \times d_f}$, $\boldsymbol{e}_u \in \boldsymbol{R}^{N \times d_u}$ and $\boldsymbol{e}_a \in \boldsymbol{R}^{N \times d_a}$ denoting the embedding representations of feature graph ($\mathcal{G}_f(\mathcal{F}, \mathcal{E}_f)$), user graph ($\mathcal{G}_u(\mathcal{U}, \mathcal{E}_u)$) and ad graph ($\mathcal{G}_a(\mathcal{A}, \mathcal{E}_a)$), respectively. $\boldsymbol{e}_{f_k} \in \boldsymbol{R}^{d_f}$ denotes the $k$-th row of $\boldsymbol{e}_f$, corresponding to $f_k$; $\boldsymbol{e}_{u_i} \in \boldsymbol{R}^{d_u}$ denotes the $i$-th row of $\boldsymbol{e}_u$, corresponding to $u_i$; and $\boldsymbol{e}_{a_j} \in \boldsymbol{R}^{d_a}$ denotes the $j$-th row of $\boldsymbol{e}_a$, corresponding to $a_j$.

## 3.4 Graph representation learning

In this section, we present graph representation learning methods for feature graph, user graph and ad graph constructed in Section 3.2, in order to enhance the performance robustness of CTR prediction. Specifically, we propose a novel causality-based feature graph representation learning method to learn graph representations of causal features, then apply GraphSAGE to learn user graph and ad graph representations.

### 3.4.1 Feature graph representation learning

We present a causality-based feature representation learning method termed GraphFwFM in gated graph neural networks (GGNNs). Simply put, GraphFwFM integrates causal inference in feature graph and the FwFMs feature interaction mechanism. GGNNs are taken as the fundamental GNNs framework because GGNNs overcome a limitation of GNNs (Li et al., 2015), i.e., no guarantee of convergence in a fixed number of steps, by using the gate mechanism in the propagation (Zhou et al., 2020). GGNNs update each node state through its neighbors and its previous time step state information using GRU. Following (Li et al., 2015), GGNNs are represented as

$$\begin{cases} \boldsymbol{H}_{f_k}^{(l)} = \left(\boldsymbol{1} - \boldsymbol{z}_{f_k}^{(l)}\right) \odot \boldsymbol{H}_{f_k}^{(l-1)} + \boldsymbol{z}_{f_k}^{(l)} \odot \overline{\boldsymbol{H}_{f_k}^{(l)}} \\ \overline{\boldsymbol{H}_{f_k}^{(l)}} = \tanh\left(\boldsymbol{W}\boldsymbol{a}_{f_k}^{(l)} + \boldsymbol{U}\left(\boldsymbol{r}_{f_k}^{(l)} \odot \boldsymbol{H}_{f_k}^{(l-1)}\right)\right) \\ \boldsymbol{z}_{f_k}^{(l)} = \sigma\left(\boldsymbol{W}_z \boldsymbol{a}_{f_k}^{(l)} + \boldsymbol{U}_z \boldsymbol{H}_{f_k}^{(l-1)}\right) \\ \boldsymbol{r}_{f_k}^{(l)} = \sigma\left(\boldsymbol{W}_r \boldsymbol{a}_{f_k}^{(l)} + \boldsymbol{U}_r \boldsymbol{H}_{f_k}^{(l-1)}\right) \\ \boldsymbol{a}_{f_k}^{(l)} = \boldsymbol{A}_{f_k:}^T \left[\boldsymbol{H}_{f_1}^{(l-1)T}, \ldots, \boldsymbol{H}_{f_S}^{(l-1)T}\right]^T + \boldsymbol{b} \\ \boldsymbol{H}_{f_k}^{(1)} = [\boldsymbol{e}_{f_k}^T, \boldsymbol{0}]^T \end{cases}, \quad (8)$$

where $\boldsymbol{A}_{f_k}$ represents the strength of node $f_k$ when it communicates with other nodes; $\boldsymbol{H}_{f_k}^{(l-1)}$ is the hidden state of node $f_k$ ($k = 1, 2, \ldots, S$) in $(l-1)$-th step; $\boldsymbol{b}$ is the bias and $\boldsymbol{e}_{f_k}$ is the initial state of node $f_k$. $\boldsymbol{z}_{f_k}^{(l)}$ and $\boldsymbol{r}_{f_k}^{(l)}$ are the update gate and the reset gate of GRU, respectively.

Each node in GGNNs has two matrices (i.e., $\boldsymbol{A}^{in}$ and $\boldsymbol{A}^{out}$) to determine the strength of information flow in and out from/to other nodes. Because each feature may behave differently when interacting with others (Ma et al., 2016; Juan et al., 2016; Juan et al., 2017), it may be insufficient to realize feature interactions in feature graph with the two interaction matrices. To this end, we design a causality-based representation learning approach (GraphFwFM) to overcome this shortcoming of GGNNs, by taking advantage of causal inference in feature graph and the FwFMs feature interaction mechanism. Figure 2 illustrates the mechanism of GraphFwFM. GraphFwFM samples the neighbors of each feature node according to the weights ($w_{k'k}^{(f)}$) of directed edges, then aggregates state information from sampled neighbors based on the principle of feature interactions in FwFMs, and updates the state information by employing the update function in GGNNs.

In GraphFwFM, each node has a state vector $\boldsymbol{H}_{f_k}^{(l)}$ which corresponds to field feature $f_k$ in the $l$-th step. Then the state of feature graph in the $l$-th step can be represented as $\boldsymbol{H}^{(l)} = [\boldsymbol{H}_{f_1}^{(l)}, \boldsymbol{H}_{f_2}^{(l)}, \ldots, \boldsymbol{H}_{f_S}^{(l)}]$. The initial state of feature graph is fed with the concatenation of field embedding ($\tilde{\boldsymbol{e}}_{f_k}$) and graph embedding ($\boldsymbol{e}_{f_k}$) of features ($\mathbb{e}_{f_k} = [\tilde{\boldsymbol{e}}_{f_k}, \boldsymbol{e}_{f_k}]$), i.e., $\boldsymbol{H}^{(0)} = [\mathbb{e}_{f_1}, \mathbb{e}_{f_2}, \ldots, \mathbb{e}_{f_S}]$. For a given node, its state is updated through the aggregation of its neighborhood nodes' states and its previous state.

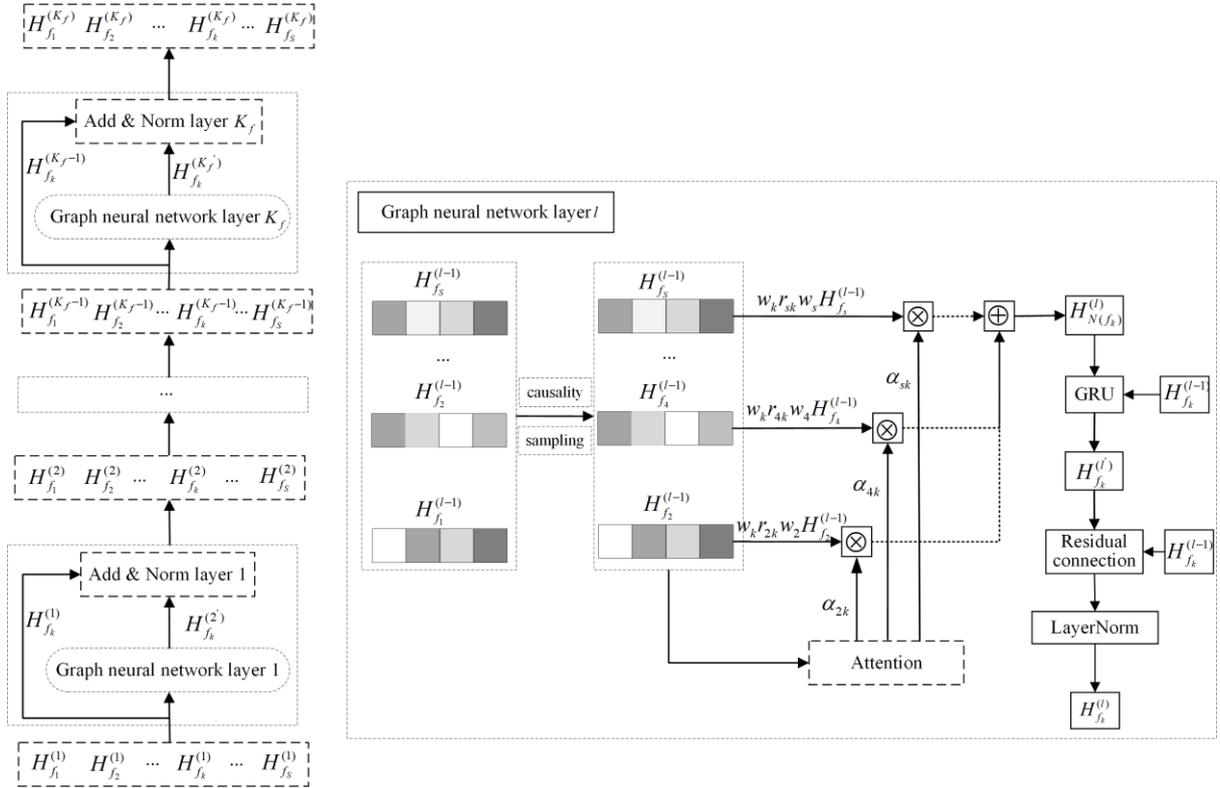

Figure 2. The mechanism of GraphFwFM for feature graph

GGNNs and other types of GNNs are prone to over-smoothing as the number of neural network layers increases (Zhou et al., 2020). As illustrated by previous research (e.g., Little & Badawy, 2019; Guo et al., 2020; Zhang et al., 2022a), sampling based on causal relationships among features can increase the performance robustness of prediction models. In this research, we sample a node's neighbors in feature graph based on the causality-based weighted adjacency matrix learnt in Section 3.2.1. In the meanwhile, for each node, the associated edges are pruned by a threshold $\varepsilon$ and the remaining neighbors form its sampled neighborhood. Formally, for node $f_k$ ($k = 1, 2, \ldots, S$),

$$w_{k'k}^{(f)} = \begin{cases} 1, & w_{k'k}^{(f)} > \varepsilon \\ 0, & w_{k'k}^{(f)} \leq \varepsilon \end{cases}, \quad k' = 1,2,\dots,S, \quad (9)$$

where $w_{k'k}^{(f)}$ is the element in the $k'$-th row and the $k$-th column of the adjacency matrix $\boldsymbol{W}^{(f)}$, which denotes whether $f_{k'}$ is a sampled neighbor of $f_k$ ($w_{k'k}^{(f)} = 1$) or not ($w_{k'k}^{(f)} = 0$).

At each step, $f_k$ interacts with other nodes in feature graph with strengths described by a transformation matrix $\boldsymbol{W}_k$. The weight of an edge from $f_{k'}$ to $f_k$ characterizes the strength of the directional interaction, i.e., $r_{k'k}$, which explicitly describes the relationship strength from $f_{k'}$ to $f_k$. At the $l$-th step, $f_k$ aggregates information of its sampled neighbors in the following way.

$$\boldsymbol{H}_{\mathcal{N}(f_k)}^{(l)} = \sum_{f_{k'} \in \mathcal{N}(f_k)} r_{k'k} \alpha_{k'k} w_{k'k}^{(f)} \boldsymbol{W}_k \boldsymbol{W}_{k'} \boldsymbol{H}_{f_{k'}}^{(l-1)}, \quad (10)$$

where $\mathcal{N}(f_k)$ is the neighborhood set of $f_k$, and $f_{k'}$ is a neighbor of $f_k$; $\boldsymbol{H}_{f_{k'}}^{(l-1)}$ is the state vector of $f_{k'}$ at the $(l-1)$-th step. $\alpha_{k'k}$ characterizes the importance of $f_{k'}$ in the aggregation for feature $f_k$.

The attention mechanism has a significant effect on the state aggregation (Veličković et al., 2017; Zhang et al., 2018; Wang et al., 2019). Thus, we design an attention-based aggregator based on $r_{k'k}$, $\boldsymbol{W}_k$ and $\boldsymbol{W}_{k'}$, which is given as

$$\alpha_{k'k} = \frac{exp(LeakyRelu(x_{k'k}))}{\sum_{f_z \in \mathcal{N}(f_k)} exp(LeakyRelu(x_{zk}))}, \quad (11)$$

$$x_{k'k} = r_{k'k} \vec{\boldsymbol{a}}^T [\boldsymbol{W}_k \boldsymbol{H}_{f_k}^{(l-1)} \parallel \boldsymbol{W}_{k'} \boldsymbol{H}_{f_{k'}}^{(l-1)}], \quad (12)$$

The new state of each feature node ($f_k$) depends on its neighbors' information $\boldsymbol{H}_{\mathcal{N}(f_k)}^{(l)}$ and its previous state $\boldsymbol{H}_{f_k}^{(l-1)}$. GRU is employed as the state update function, in order to improve the long-term propagation across graph structure and deepen GraphFwFM. The operation $Add\&Norm$ in transformer has an excellent performance in prediction (Vaswani et al., 2017). In particular, the $Add$ operation applies a residual connection between each pair of two neural network layers, which combines low-order and high-order features; and the $Norm$ operation employs the layer normalization function to stabilize each hidden layer (Ba et al., 2016). The state update can be formally expressed as follows.

$$\boldsymbol{H}_{f_k}^{(l')} = GRU\big(\boldsymbol{H}_{\mathcal{N}(f_k)}^{(l)}, \boldsymbol{H}_{f_k}^{(l-1)}\big), \quad (13)$$

$$\boldsymbol{H}_{f_k}^{(l)} = LayerNorm(\boldsymbol{H}_{f_k}^{(l')} + \boldsymbol{H}_{f_k}^{(l-1)}), \quad l = 0,1,\dots,K_f, \quad (14)$$

where $K_f$ is the depth of GraphFwFM for feature graph.

### 3.4.2 User graph and ad graph representation learning

In user (and ad) graph, when two users (and ads) are connected, their neighborhoods might be similar. In this research, considering the neighborhood similarity of connected users (and ads), we employ GraphSAGE to capture graph representations of users and ads are learnt by using relationships between nodes in user graph and ad graph.

GraphSAGE is an inductive graph representation learning approach that can generate node representation for previously unseen nodes (Hamilton et al., 2017). In order to generalize to unseen nodes, GraphSAGE learns a function to generate a node's representation by sampling and aggregating its local neighborhood. In user graph, user representation can be learnt through the following neural network

$$\begin{cases} \boldsymbol{H}_{\mathcal{N}(u_i)}^{(l)} = Aggregator(\boldsymbol{H}_u^{(l-1)}, \forall u \in \mathcal{N}(u_i)) \\ \boldsymbol{H}_{u_i}^{(l)} = \sigma\left(\boldsymbol{W}^k \cdot concat(\boldsymbol{H}_{u_i}^{(l-1)}, \boldsymbol{H}_{\mathcal{N}(u_i)}^{(l)})\right) \end{cases}, \quad l = 1,2,\dots,K_u, \quad (15)$$

where $Aggregator$ denotes the aggregator function, $l$ is the index of a GraphSAGE layer, $K_u$ is the depth of GraphSAGE in user graph, $\mathcal{N}(u_i)$ is the sampled neighborhood of node $u_i$ based on the weights of edges computed in Section 3.2.2, and $\boldsymbol{h}_{u_i}^0 = \boldsymbol{e}_{u_i}(i = 1,2,\dots,N)$.

Similarly, graph representation of ad graph can also be obtained by using GraphSAGE, denoted as $\boldsymbol{H}_{a_j}^{(l)}(l = 1,2,\dots,K_a, j = 1,2,\dots,M)$.

### 3.5 Attention

This research applies the multi-head self-attention mechanism to discriminate the importance of different neural network layers. The multi-head self-attention mechanism is the core idea of the transformer, which maps a query and set of key-value pairs with multiple different linear projections and jointly integrates these sub-space representations by a concatenation operation to capture information from various sources (Vaswani et al., 2017; Liang et al., 2022). The multi-head self-attention can be formulated as

$$\begin{cases} H_h = softmax\left(\frac{Q_h K_h^T}{\sqrt{d_K}}\right) V_h \\ Q_h = Q W_h^{(Q)}, K_h = K W_h^{(k)}, V_h = V W_h^{(v)} \end{cases}, \quad (16)$$

$$H = [H_1, H_2, \ldots, H_\hbar] W^0, \quad (17)$$

where $W_h^{(Q)} \in \mathbb{R}^{n \times d_K}$, $W_h^{(k)} \in \mathbb{R}^{n \times d_K}$, $W_h^{(v)} \in \mathbb{R}^{n \times d_V}$ are parameter matrices learnt from the $h$-th head, $h = 1,2,\ldots,\hbar$, $d_K$ is the dimension of queries and keys, $d_V$ is the dimension of values, $W^0 \in \mathbb{R}^{(\hbar \times d_V) \times n}$ is a parameter matrix projecting the concatenation of $\hbar$ heads into the output space $\mathbb{R}^n$, and $H_h$ is the output matrix of the $h$-th head.

In each head, the attention score is normalized by the softmax function (Equation 16). The multi-head self-attention mechanism concatenates results from multiple heads to obtain the final representation (Equation 17).

In feature graph representation learning described in Section 3.4.1, graph state at the $l$-th step is obtained from the $l$-th layer of GraphFwFM, i.e., $H^{(l)} = \left[H_{f_1}^{(l)}, H_{f_2}^{(l)}, \ldots, H_{f_S}^{(l)}\right]$, ($l = 0, 1, 2, \ldots, K_f$). Conceptually, GraphFwFM with $K_f$ layers produces multiple orders of graph representations. Following prior research (e.g., He et al., 2021; Liu et al., 2021), let $Q = K = V = \left[H^{(0)}, H^{(1)}, \ldots, H^{(K_f)}\right]^T$. This research utilizes the multi-head self-attention mechanism to learn the effect of low-order and high-order features simultaneously (Gao et al., 2018), which is given as

$$[H_{f_1}, H_{f_2}, \ldots, H_{f_S}] = multihead\left(\left[H^{(0)}, H^{(1)}, \ldots, H^{(K_f)}\right]^T\right), \quad (18)$$

where $H_{f_k}$ is the resulting node representation of $f_k$ in feature graph.

In order to accurately characterize the importance of users and ads, the multi-head self-attention mechanism allocates a specific weight to each user and each ad. Specially, we apply multi-head self-attention mechanism on the representation of each user ($\left[H_{u_i}^{(0)}, H_{u_i}^{(1)}, \ldots, H_{u_i}^{(K_u)}\right]$, $i = 1,2,\ldots N$) in user graph and the representation of each ad ($\left[H_{a_j}^{(0)}, H_{a_j}^{(1)}, \ldots, H_{a_j}^{(K_a)}\right]$, $j = 1,2,\ldots M$) in ad graph, respectively. In a similar way, we can obtain new user representation $H_{u_i}$ ($i = 1,2,\ldots N$) and ad representation $H_{a_j}$ ($j = 1,2,\ldots M$) through attention mechanisms.

## 3.6 Prediction

In the prediction layer, this research takes a simple neural-based attention prediction method to determine contributions of graph representations of features, users and ads in predicting the CTR (Li et al., 2020).

Specifically, we utilize a fully connected layer for each graph representation and then perform a weighted sum over these representations, which is given as

$$att_{u_i} = tanh(W_u H_{u_i} + b_u),$$

$$att_{a_j} = tanh(W_a H_{a_j} + b_a),$$

$$att_{f_k} = tanh(W_{f_k} H_{f_k} + b_{f_k}), \quad (19)$$

$$H = att_{u_i} \odot H_{u_i} + att_{a_j} \odot H_{a_j} + \sum_k att_{f_k} \odot H_{f_k}, \quad (20)$$

where $W_u$, $W_a$ are weight matrices for user graph and ad graph, respectively, $W_{f_k}$ is weight matrix for the $k$-th feature ($f_k$) in a fully connected neural network, $b_u$, $b_a$ and $b_{f_k}$ are bias vectors for users, ads and feature $f_k$ ($k = 1,2,..,S$), respectively, $\odot$ is the element-wise multiplication, $att_{u_i}$, $att_{a_j}$ and $att_{f_k}$ are weight vectors for user $u_i$, ad $a_j$ and feature $f_k$, respectively.

The vector $H$ is taken as the input of a fully connected neural network layer to make CTR prediction where the sigmoid function squeezes predicted values to [0,1], given as

$$\hat{y} = sigmoid(WH + b). \quad (21)$$

## 4. Experiments

This section starts with experimental settings including datasets, evaluation metrics, then turns to model comparison, and finally makes hyper-parameter analysis and ablation study of the proposed model.

### 4.1 Datasets

We evaluate the proposed model (Causal-GNN) and baselines on the following three benchmark datasets that have been widely used in the extant literature on CTR prediction[1].

- **Criteo**[2]: The Criteo display advertising challenge dataset is provided by CriteoLab on Kaggle in 2014. Each record includes 13 numerical fields and 26 categorical fields.
- **Avazu**[3]: The Avazu dataset is with the Kaggle 2014 CTR prediction competition, containing Avazu data during 11 days, ordered chronologically. Each record includes 23 fields such as ad-id, site-id, etc.
- **MovieLens-1M**[4]: The MovieLens-1M dataset is collected from MovieLens Web site, which contains 1,000,209 ratings from 6,040 anonymous users on about 3,900 movies. Each record includes user-id, ad-id, rating, timestamp, gender, age, occupation, Zip-code, title, and genres.

We randomly sample 2 million instances in Criteo, Avazu, and use all instances in MovieLens-1M for experimental evaluation. Each dataset is divided into three parts - training (80%), validation (10%) and test (10%).

## 4.2 Evaluation metrics

This study employs the two most popular metrics, namely AUC-ROC and Logloss, to evaluate the performance of Causal-GNN and baselines in CTR prediction tasks.

- **AUC-ROC** is defined as the area under the receiver operating characteristics (ROC) curve, measuring the probability of positive instances ranked higher than negative ones. The ROC curve is based on the ratio of true positive rate (TPR) and false positive rate (FPR). A higher AUC-ROC indicates the better prediction performance.
- **Logloss** reflects the average deviation of predicted values from true values. A lower Logloss suggests the better prediction capacity. Specifically, the Logloss is

---

[1] Note that MovieLens-1M is a popular dataset for CTR prediction in recommender systems that are a similar Web service to online advertising (Yang and Zhai, 2022). This study chooses MovieLens-1M due to the following reasons: (1) it provides rich information for evaluating the strength of modeling components in Causal-GNN; (2) it can be used to illustrate the generalization ability of Causal-GNN in a different but related application.

[2] https://www.kaggle.com/c/criteo-display-ad-challenge

[3] https://www.kaggle.com/c/avazu-ctr-prediction/data

[4] https://grouplens.org/datasets/movielens/

formulated as

$$L = -\frac{1}{N_s}\sum_{i=1}^{N_s}[y_i \times log\hat{y}_i + (1 - y_i)(1 - log\hat{y}_i)], \quad (22)$$

where $y_i = 0$ (or 1) denotes the label, $\hat{y}_i \in [0,1]$ is the predicted CTR, $N_s$ is the size of the dataset.

## 4.3 Experimental settings

### 4.3.1 Baselines

This study validates the performance of Causal-GNN by comparing with a class of state-of-the-art models, including low-order models (i.e., LR, FMs, AFMs), high-order models (i.e., NFMs, CIN, DeepFM) and graph models (i.e., Fi-GNN, GAT, GraphFM), as specified below.

- **Logistic regression (LR)** is one of the most widely used baseline method, which is a linear combination of individual features.
- **Factorization Machines (FMs)** (Rendle, 2010) is a second-order feature interaction model, supporting the better prediction on sparse data by using factorized parameters.
- **Attentional factorization machine (AFM)** (Xiao et al., 2017) utilizes the attention-based pooling mechanism to represent contributions of various feature interactions.
- **Neural factorization machine (NFM)** (He & Chua, 2017) is an implicit high-order feature interaction model, which describes second-order feature interactions by using FMs, and captures high-order feature interactions through a DNNs-based component.
- **Compressed Interaction Network (CIN)** (Lian et al., 2018) is an explicit high-order model, which captures feature interactions at the vector-wise level and compresses the intermediate tensor to update the feature map.
- **DeepFM** (Guo et al., 2017) is an end-to-end model, which integrates FMs to capture low-order feature interactions and DNNs to capture high-order feature interactions.
- **Feature interaction graph neural networks (Fi-GNN)** (Li et al., 2019b) is a GNNs-based feature interaction model, which realizes field-aware feature interactions in a graph structure.
- **Graph attention networks (GAT)** (Veličković et al., 2017) utilizes attention

mechanisms to assign the importance to neighbors of a focal node, with no requirement of the prior knowledge of the entire graph structure.

- **Graph Factorization Machine (GraphFM)** (Li et al., 2021b) integrates the interaction function of FMs into the feature aggregation strategy of GNNs and captures higher-order feature interactions with the increase of stacking layers.

### 4.3.2 Parameter settings

In the following experiments, for the purpose of fair comparison, the Logloss (Equation 22) is used as the objective function for all the prediction methods, Adam is taken as the optimizer, and batch size is set as 512.

In order to achieve stable results, for each method, we perform three times of experiments, and in each experiment 12 epochs were run to reduce the influence of random factors. The early stopping strategy is used to interrupt the training when the Logloss on the validation set does not decrease in five successive epochs. Final results were calculated as the average of the three experimental results.

In order to capture the local optimal result, we use a grid search method to determine hyper-parameters of the proposed model (Causal-GNN). Specifically, the learning rate is searched in the interval [0.0005, 0.001, 0.0015, 0.002, 0.0025, 0.003], and we obtained different values for the three datasets, i.e., 0.003 on Criteo, 0.002 on Avazu, and 0.0015 on MovieLens-1M. Similarly, the embedding size in feature graph is searched in the interval [8, 16, 32, 64, 128], and we took the value of 128 for the three datasets. The embedding sizes in user graph and ad graph are both set as 64. The number of GNNs layers is set as 3. Moreover, on MovieLens-1M, the aggregator used in GraphSAGE dealing with user graph and ad graph is chosen from four candidates namely mean aggregator, LSTM aggregator, mean pooling aggregator and max pooling aggregator. We found that max pooling aggregator achieves the best performance.

For fair comparison, we also set the number of DNNs and GNNs layers of the baselines (i.e., NFM and DeepFM) as 3. Meanwhile, as for other settings of the baselines, we use parameters suggested in articles originating these methods to ensure the best performance of these methods.

## 4.4 Performance comparison

In the following, we report the performance comparison among Causal-GNN and baselines for CTR prediction tasks on three datasets. Table 2 presents the performance comparison of these models.

Table 2. Performance comparison

| Model | Criteo | | Avazu | | MovieLens-1M | |
|---|---|---|---|---|---|---|
| | AUC | Logloss | AUC | Logloss | AUC | Logloss |
| LR | 0.7652 | 0.4782 | 0.7626 | 0.3800 | 0.8219 | 0.3426 |
| FMs | 0.7502 | 0.4850 | 0.7699 | 0.3765 | 0.8243 | 0.3405 |
| AFMs | 0.7544 | 0.4820 | 0.7667 | 0.3780 | 0.8238 | 0.3415 |
| NFM | 0.7561 | 0.4807 | 0.7694 | 0.3764 | 0.8339 | 0.3349 |
| CIN | 0.7663 | 0.4753 | 0.7614 | 0.3865 | 0.8280 | 0.3372 |
| DeepFM | 0.7787 | 0.4645 | 0.7666 | 0.3793 | 0.8286 | 0.3359 |
| Fi-GNN | 0.7839 | 0.4609 | 0.7684 | 0.3761 | 0.7840 | 0.3648 |
| GAT | 0.7726 | 0.4698 | 0.7644 | 0.3785 | 0.7271 | 0.3955 |
| GraphFM | 0.7837 | 0.4601 | **0.7741** | 0.3772 | 0.7843 | 0.3642 |
| Causal-GNN | **0.7844** | **0.4599** | 0.7728 | **0.3750** | **0.8342** | **0.3298** |

From Table 2, we can observe the following results. First, overall our proposed method (Causal-GNN) outperforms the baselines in terms of AUC and Logloss on the three datasets, with the exception that GraphFM outperforms Causal-GNN in AUC on Avazu at the 0.001 level. Specifically, on the three datasets, compared with the baselines except for GraphFM, the CTR prediction performance in AUC is improved by Causal-GNN significantly at the 0.01 level; in the meanwhile, Causal-GNN achieves smaller Logloss at the 0.001 level than the baselines.

Second, it is worthwhile to note that MovieLens-1M contains specific information of users and ads, while Criteo and Avazu do not. Thereby, MovieLens-1M supports the full Causal-GNN model validation. Thus, the superiority of Causal-GNN on MovieLens-1M illustrates the effectiveness of the Causal-GNN model, while its superiority on Criteo and Avazu proves the effectiveness of the GraphFwFM on the causal feature graph.

Third, Causal-GNN overwhelmingly outperforms the low-order models (i.e., LR, FMs, AFMs). This indicates that GNNs-based feature representations entitle Causal-GNN to capture more sophisticated features.

Fourth, Causal-GNN has comparable performance to high-order models (NFMs, CIN, DeepFM). In this sense, Causal-GNN can retain better interpretability of feature interactions with no sacrifice of performance.

Fifth, the superiority of Causal-GNN over graph models (i.e., Fi-GNN, GraphFM and GAT) can be attributed to the effectiveness of causal inference among features and graph representation learning on feature graph, user graph and ad graph. Note that experiments are conducted using the original Fi-GNN and GraphFM codes with the same settings of parameters specified by Li et al. (2019b) and Li et al. (2021b). We can notice that the performance results of Fi-GNN and GraphFM illustrated in Table 2 differ from those reported in Li et al. (2019b) and Li et al. (2021b). Because the codes and experimental settings are exactly the same, we speculate that, (1) the performance difference of Fi-GNN and GraphFM on Criteo and Avazu may result from the fact that Li et al. (2019b) and Li et al. (2021b) used all instances in Criteo and Avazu, while this research used 2 million sampled instances to conduct experiments; (2) as for the performance difference of Fi-GNN and GraphFM on MovieLens-1M, it may be imputed to that the way that Li et al. (2021b) converted the label values and removed some instances in MovieLens-1M. MovieLens-1M (about 1,000,000 instances) has 5 levels of ratings (1-5) as labels. These labels need to be converted into 2 labels (click or not click). As described in Li et al. (2021b), instances with labels 1-2 and those with labels 4-5 were treated as negative samples (not click) and positive samples (click), respectively, and removed neutral instances with ratings of 3 (about 200,000 instances), which may most likely lead to prediction biases. However, in order to get valid results, this research retains all instances in MovieLens-1M to conduct experiments.

Last but not the least, the superiority of Causal-GNN on cross-domain CTR prediction tasks, i.e., display ads (Criteo), mobile ads (Avazu) and movie recommendations (MovieLens-1M) may reveal its good generalization ability. We will explore this issue in more detail in

Section 4.6.3.

## 4.5 Hyper-parameter analysis

In order to get deep insights into the architecture of the proposed model (Causal-GNN), we study effects of hyper-parameters including the learning rate, the embedding size in GraphFwFM and different aggregators in GraphSAGE, on the prediction performance.

### 4.5.1 Effect of learning rate

The learning rate determines the learning speed of a supervised learning model and influences how and when the objective function converges to the local optimum. In the experiment, we search the learning rate in the interval [0.0005, 0.001, 0.0015, 0.002, 0.0025, 0.0030] for Causal-GNN on the three datasets. Figure 3 presents the performance of Causal-GNN with different learning rates.

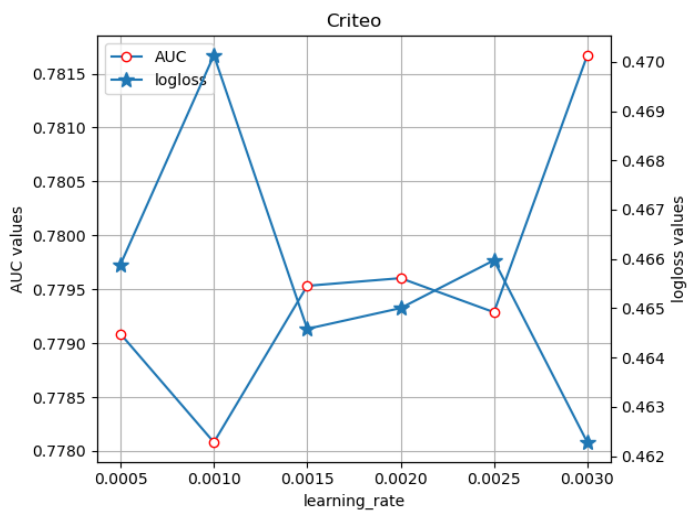

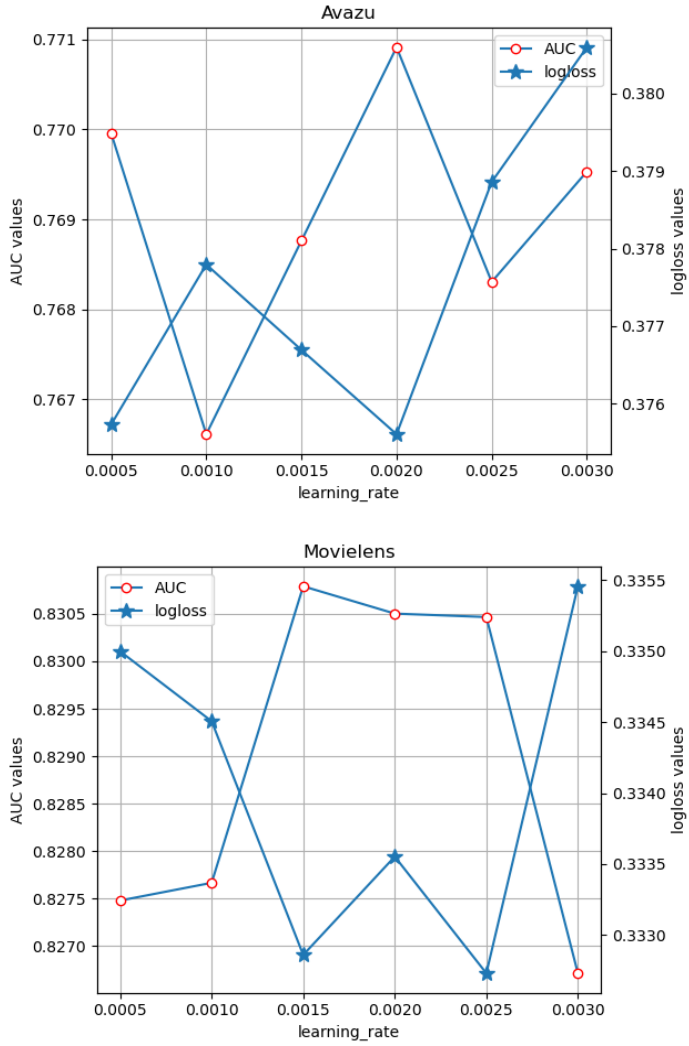

Figure 3. The performance of Causal-GNN with different learning rates

From Figure 3, we can observe that AUC and Logloss change in converse directions with the increase of the learning rate. In other words, the performance of Causal-GNN is consistent under the two evaluation metrics. Moreover, the local optimal learning rate is different on the three datasets: 0.003 on Criteo, 0.002 on Avazu and 0.0015 on MovieLens-1M. That is, Causal-GNN performs at different convergence speeds on the three datasets, possibly due to the difference in data characteristics. The three datasets contain different numbers of fields: Criteo has 39 fields, Avazu has 23 fields and MovieLens-1M has 10 fields. Intuitively, a richer dataset may require a larger learning rate for a given model to achieve the local convergence. This may explain the fact that Causal-GNN is assigned the largest learning rate (0.003) on Criteo and the smallest learning rate (0.0015) on MovieLens-1M.

**4.5.2 Effect of embedding size**

The embedding size is a significant parameter in deep learning frameworks, which substantially influences the model performance and the computational cost. We explore the embedding size with a grid search method for Causal-GNN on the three datasets. Figure 4 presents the performance of Causal-GNN with different embedding sizes in feature graph.

    The latent representation preserves more information with the increase of the embedding size. From Figure 4, we can see that the embedding size is a sensitive parameter in feature representations of Causal-GNN. The performance of Causal-GNN becomes better when the embedding size increases, although there is a small-range fluctuation on Avazu. Results show that the embedding size is optimal at 128 on all three datasets. This may indicate that the three datasets require strong representations to fit the data.

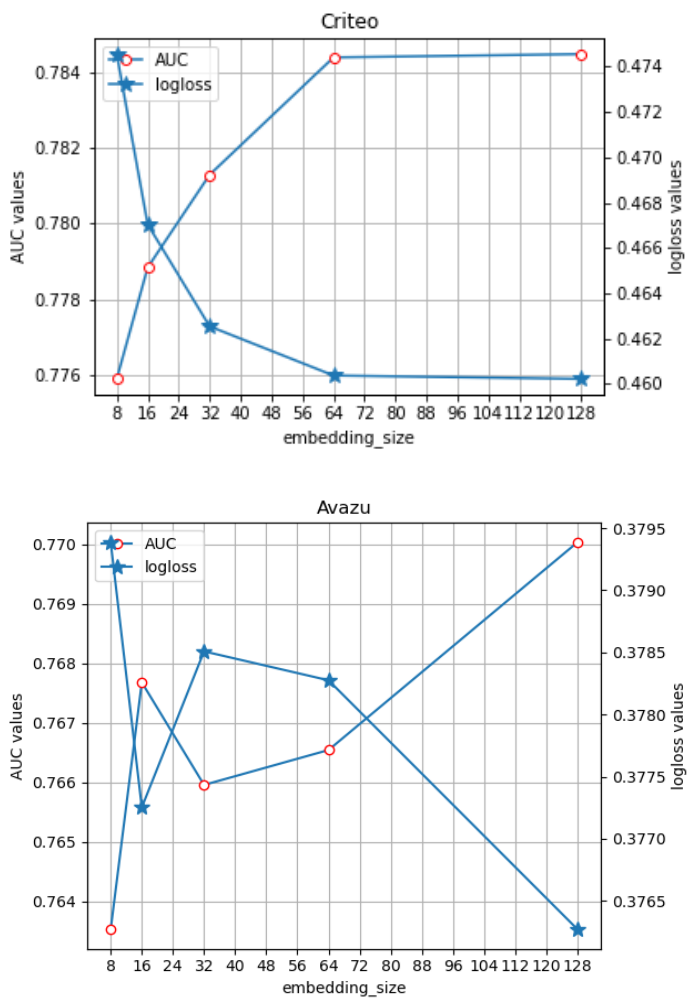

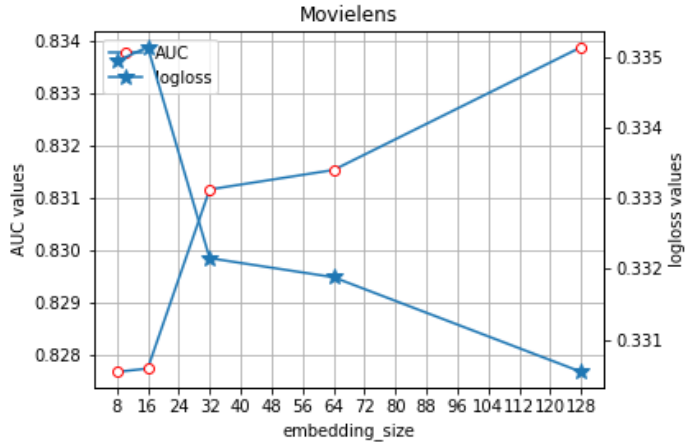

Figure 4. The performance of Causal-GNN with different embedding sizes

### 4.5.3 Effect of aggregator

Aggregator is an important operation in GraphSAGE. An ideal aggregator should be symmetric and trainable. Hence, choosing a good aggregator can improve the representation capacity of GNNs. Among the three datasets, only MovieLens-1M provides necessary information to build user graph and ad graph. Therefore, this study explores effects of four aggregators, namely mean aggregator, LSTM aggregator, mean pooling aggregator and max pooling aggregator, on the performance of Causal-GNN on MovieLens-1M, as shown in Figure 5.

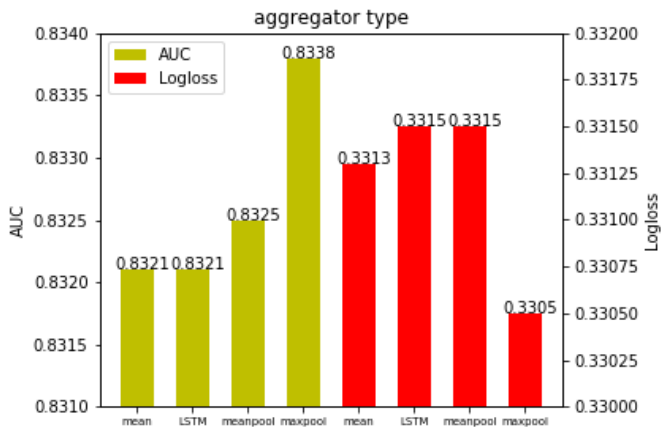

Figure 5. The performance of Causal-GNN with different aggregators on MovieLens-1M

From Figure 5, we can observe that the max pooling aggregator results in the best AUC performance of Causal-GNN while the mean aggregator and the LSTM aggregator lead to the worst AUC. Note that the mean aggregator and the LSTM aggregator have a small difference in AUC at the 0.00001 level. From the perspective of Logloss, the max pooling aggregator results in the smallest Logloss performance of Causal-GNN, while the LSTM aggregator has the largest Logloss. Thus, we can conclude that the max pooling aggregator is the best aggregator and the LSTM aggregator is the worst aggregator on MovieLens-1M. The possible explanation is that the LSTM aggregator is not inherently symmetric in that it processes the inputs as a sequence (Hamilton et al., 2017). Additionally, the mean aggregator has an unsatisfactory performance because it is untrainable, whereas a trainable aggregator is indispensable for GNNs-based models.

## 4.6 Ablation study

We conduct ablation studies to evaluate the contribution of modeling components of Causal-GNN. Specifically, we remove the GraphFwFM to valid the effect of our proposed GGNNs-based graph representation learning approach in Section 4.6.1, remove each of the three graph representations to test the effect of graph representations of features, users and ads in Section 4.6.2, and remove the component of causal learning on feature graph to examine the role of causality in feature representations in Section 4.6.3.

### 4.6.1 Effect of GraphFwFM

We examine how our proposed graph representation learning component (GraphFwFM) on feature graph benefits Causal-GNN. In particular, we remove the GraphFwFM component (i.e., model-GraphFwFM) and compare its performance with the full Causal-GNN model, as illustrated in Table 3. From Table 3, we can notice that, the model without GraphFwFM performs poorly comparing with the full model. This indicates that GraphFwFM is substantially effective in learning graph representations of features.

Table 3. Effect of GraphFwFM

|  | Criteo | | Avazu | | MovieLens-1M | |
| --- | --- | --- | --- | --- | --- | --- |
|  | AUC | Logloss | AUC | Logloss | AUC | Logloss |
| Model-GraphFwFM | 0.7722 | 0.4694 | 0.7637 | 0.3814 | 0.7591 | 0.4601 |
| Causal-GNN model | 0.7844 | 0.4599 | 0.7728 | 0.3750 | 0.8342 | 0.3298 |

### 4.6.2 Effect of graph representations of feature, user and ad

We examine how Causal-GNN is benefited from graph representations of features, users and ads, respectively. Specifically, we remove each of the three graph representations and compare these models with the full Causal-GNN model, as shown in Table 4.

Table 4. Effect of graph representations of feature, user and ad on MovieLens-1M

| Model | AUC | Logloss |
| --- | --- | --- |
| User graph representation | 0.7232 | 0.3960 |
| Ad graph representation | 0.7639 | 0.3756 |
| Feature graph representation | 0.7918 | 0.3584 |
| User+ad graph representations | 0.8273 | 0.3350 |
| User+feature graph representations | 0.8042 | 0.3517 |
| Ad+feature graph representations | 0.8287 | 0.3324 |
| Causal-GNN model | 0.8342 | 0.3298 |

From Table 4, we can observe that removing one or more graph representations leads to the worse performance than the full Causal-GNN model. Moreover, among the three graph representations, feature graph is the most effective in improving the prediction performance. Among combinations of two graph representations, user+ad and ad+feature graph have comparable performance, and user+feature graph leads to the worst performance, which suggests that ad graph is an important input for CTR prediction.

### 4.6.3 Effect of causal inference in feature representation

In the following we examine the effect of causal inference in feature representation. First, we explore how causal inference among features improves the performance of Causal-GNN. Specifically, we replace causal feature graph with a complete feature graph (i.e., model-causality) and compare its performance with the full Causal-GNN model, as demonstrated in Table 5.

From Table 5, we can see that model-causality performs worse than the full Causal-GNN model. Moreover, it is apparent that the model performance on Criteo is the most affected by the causality among those on the three datasets. The possible reason is that Criteo has more fields and in turn possesses richer causal relationships among features than the other two datasets.

Table 5. Effect of causal inference on the performance

|  | Criteo | | Avazu | | MovieLens-1M | |
| --- | --- | --- | --- | --- | --- | --- |
|  | AUC | Logloss | AUC | Logloss | AUC | Logloss |
| Model-causality | 0.7796 | 0.4650 | 0.7674 | 0.3772 | 0.8285 | 0.3342 |
| Causal-GNN model | 0.7844 | 0.4599 | 0.7728 | 0.3750 | 0.8342 | 0.3298 |

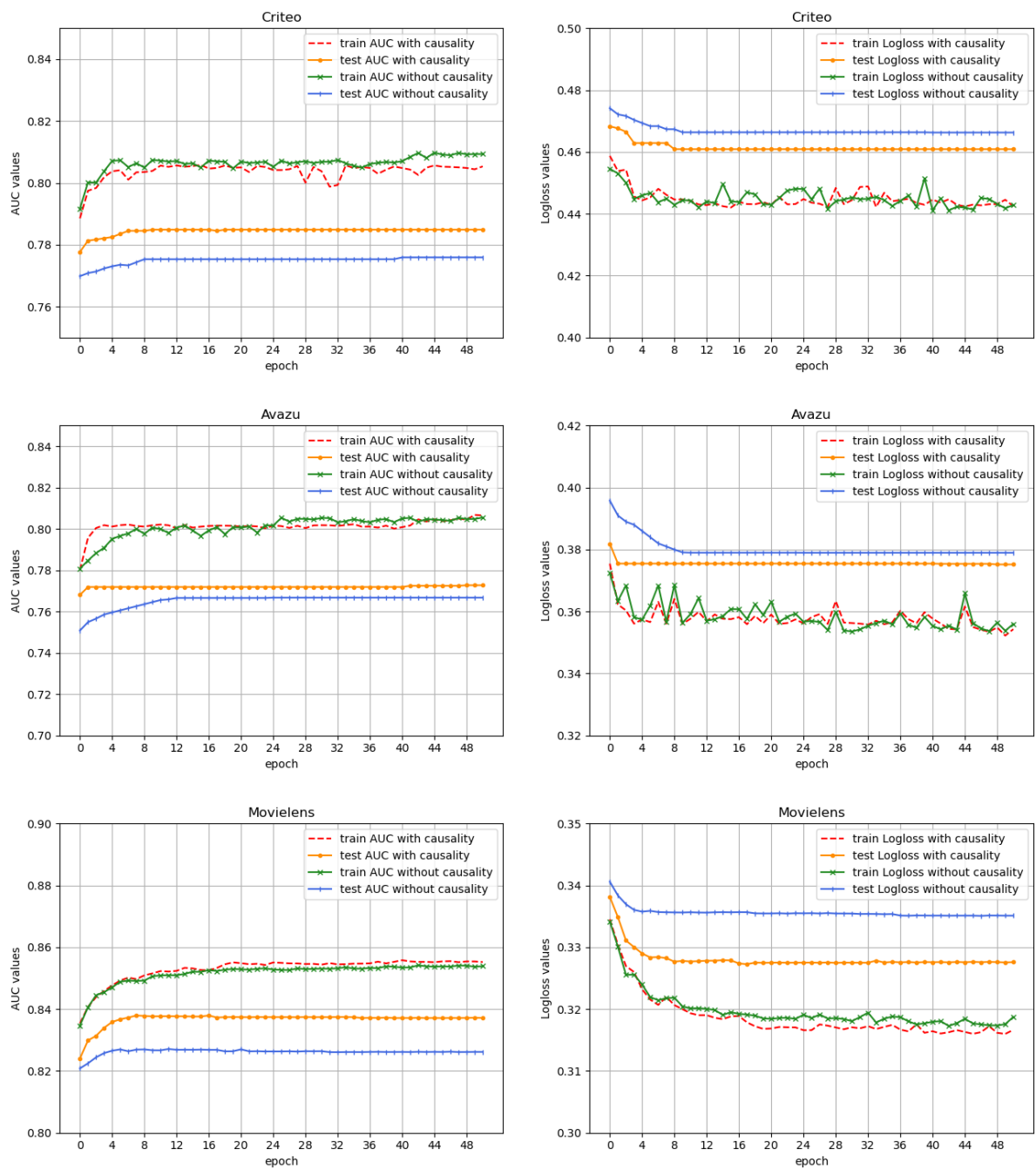

Figure 6. The model performance in the training phase and the testing phase

Next, we inspect how causal inference among features enhances the generalization of causal-GNN. Specifically, we compare the reduced model without causality (model-causality) with the full Causal-GNN model with respect to how the model performance evolves in the training phase and the testing phase when the number of epochs increases, as demonstrated in Figure 6.

From Figure 6, we can see that the performance difference between the training phase and the testing phase achieved by the full Causal-GNN model is much smaller than that by its reduced form without causality, in terms of both AUC and Logloss on the three datasets. This implies that causal inference among features makes Causal-GNN less prone to over-fitting. In other words, Causal-GNN learns underlying rules and patterns in the data, rather than shallow information. Moreover, compared to the reduced form without causality, the full Causal-GNN model can reach the stable performance faster when the number of epochs increases both in the training phase and in the testing phase.

## 5. Conclusion

In this study, we propose a causality-based CTR prediction model in the GNNs framework (Causal-GNN) integrating multiple representations (feature graph, user graph and ad graph). Moreover, we proposed a graph representation learning approach (GraphFwFM) based on GGNNs to capture causal relationships in feature graph. GraphSAGE is employed to obtain graph representations of users and ads. Experiments conducted on three public datasets (Criteo, Avazu and MovieLens-1M) show that Causal-GNN achieves better results than state-of-the-art baselines for CTR prediction tasks and GraphFwFM is substantially effective in capturing high-order sophisticated representations in causal feature graph.

This research also yields several interesting findings that provide valuable managerial insights for online advertisers and platform providers. First, causal relationships among features are essential information for CTR prediction. Hence, it is important to explicitly derive causal relationships either manually or statistically and focus on roles of casual factors in generating more clicks. We believe that empirical studies on causal relationships among various advertising variables certainly deserve further investigation. Note that this research focuses on the development of causal learning based GNNs models for CTR prediction; in the meanwhile, the benchmark datasets used in the experiments are for prediction tasks, rather than empirical studies. Second, user graph and ad graph can enhance CTR prediction. This implies that relationships among users and among advertisements complement to users' and advertisements' characteristics in determining the CTR. This reminds advertisers to carefully design advertising portfolios and inspect relationships among their advertisements. Moreover, it also implies that whether a specific user will click an advertisement may be influenced by their friends, which is consistent with simulation results provided by Yang et al. (2018). In this sense, both advertisers and online platforms should try all means to improve interactions among users (Yang & Gao, 2021). Third, the comparison experiment illustrates that graph models are generally superior to non-graph ones, which indicates that it is promising to handle the CTR prediction problem in a graph structure.

In the future research, we plan to explore more sophisticated feature interactions in the GNNs framework. That is, it aims to study flexible causality-based interaction rules and diverse aggregate strategies in graph neural architectures in order to achieve the better performance for CTR prediction. Second, how to build interpretable GNNs-based CTR prediction models is a meaningful research perspective. In particular, we are intended to integrate the domain-specific knowledge with causal inference to support explicit representations of high-order feature aggregations and interactions. Then functionally-grounded evaluation metrics of model interpretability can be developed for prediction tasks. Last but not the least, it is desirous to build real-time prediction models through designing efficient computing architectures and search strategies.

## Acknowledgements

We are thankful to the editor and anonymous reviewers who provided valuable suggestions that led to a considerable improvement in the organization and presentation of this manuscript. This work is partially supported by the (NSFC National Natural Science Foundation of China) grants (72171093, 62072026) and Beijing Natural Science Foundation grant (JQ20022).